\documentclass[conference]{IEEEtran}
\IEEEoverridecommandlockouts

\usepackage{cite}
\usepackage{amsmath,amssymb,amsfonts}
\usepackage{algorithmic}
\usepackage{graphicx}
\usepackage{textcomp}
\usepackage{xcolor}
\def\BibTeX{{\rm B\kern-.05em{\sc i\kern-.025em b}\kern-.08em
    T\kern-.1667em\lower.7ex\hbox{E}\kern-.125emX}}

\usepackage{url}
\usepackage{longtable}
\usepackage{booktabs}
\usepackage{multicol}
\usepackage{multirow}
\usepackage{lineno}
\usepackage{xspace}
\usepackage{multirow}
\usepackage{textcomp}
\usepackage{graphicx}
\usepackage{caption}
\usepackage{subcaption}
\usepackage{float}
\usepackage{hyperref}  
\usepackage[colorinlistoftodos]{todonotes}

\newtheorem{definition}{Definition}[section]

\newcommand{\eg}{\emph{e.g.}}
\newcommand{\ea}{\emph{et al.}}

\begin{document}

\title{Spike2Vec: An Efficient and Scalable Embedding Approach for
  COVID-19 Spike Sequences}

\author{\IEEEauthorblockN{1\textsuperscript{st} Sarwan Ali}
\IEEEauthorblockA{\textit{Department of Computer Science} \\
\textit{Georgia State University}\\
Atlanta, GA, USA \\
sali85@student.gsu.edu}
\and
\IEEEauthorblockN{2\textsuperscript{nd} Murray Patterson}
\IEEEauthorblockA{\textit{Department of Computer Science} \\
\textit{Georgia State University}\\
Atlanta, GA, USA \\
mpatterson30@gsu.edu}
}

\maketitle

\begin{abstract}

With the rapid global spread of COVID-19, more and more data related
to this virus is becoming available, including genomic sequence data.
The total number of genomic sequences that are publicly available on
platforms such as GISAID is currently several million, and is
increasing with every day.  The availability of such \emph{Big Data}
creates a new opportunity for researchers to study this virus in
detail.  This is particularly important with all of the dynamics of
the COVID-19 variants which emerge and circulate.  This rich data
source will give us insights on the best ways to perform genomic
surveillance for this and future pandemic threats, with the ultimate
goal of mitigating or eliminating such threats. Analyzing and
processing the several million genomic sequences is a challenging
task.  Although traditional methods for sequence classification are
proven to be effective, they are not designed to deal with these
specific types of genomic sequences.  Moreover, most of the existing
methods also face the issue of scalability. Previous studies which
were tailored to coronavirus genomic data proposed to use spike
sequences (corresponding to a subsequence of the genome), rather than
using the complete genomic sequence, to perform different machine
learning (ML) tasks such as classification and clustering.  However,
those methods suffer from scalability issues.

In this paper, we propose an approach called Spike2Vec, an efficient
and scalable feature vector representation for each spike sequence
that can be used for downstream ML tasks.  Through experiments, we
show that Spike2Vec is not only scalable on several million spike
sequences, but also outperforms the baseline models in terms of
prediction accuracy, F1 score, etc.  Since this type of study on such
huge numbers of spike sequences has not been done before (to the best
of our knowledge), we believe that it will open new doors for
researchers to use this data and perform different tasks to unfold new
information that was not available before.  We also
use information gain (IG) to compute the importance of each amino acid
in the spike sequence.  The amino acids with higher IG values tend to
be the same as many reported by the USA based Centers for Disease
Control and Prevention (CDC) for different variants.

\end{abstract}


\section{Introduction}

Very few fields of study remain untouched in the big data era, as
massive amounts of data are collected in every domain from
finance~\cite{sharma2019financial,dong2020belt} to
astronomy~\cite{kremer2017astronomy,mace2020neowise}.  The field of
biomedical and health informatics is no exception; one which has had a
recent and rather rapid growth spurt in the amount of available data,
due to the COVID-19
pandemic~\cite{leung2020olap,leung2020science,ali2021efficient,tayebi2021robust}.
One facet of this increase is the amount of genomic data becoming
available for COVID-19 in databases such as
GISAID~\cite{gisaid_website_url}, where several million viral genome
(virome) sequences of COVID-19 --- or more precisely,
SARS-CoV-2\footnote{Severe Acute Respiratory Syndrome Coronavirus 2
  (SARS-CoV-2), the virus which causes the COVID-19 disease} --- are
available.

Such data has a high \emph{volume}, as the SARS-CoV-2 virome has
${\approx}30$K nucleotide base-pairs, and there are more than 2.5
million such sequences available in GISAID alone.  While the number
COVID-19 patients being sequenced is a fraction of the actual number
of cases, the sheer number of infections (both now and in the past)
means that the \emph{velocity} in which SARS-CoV-2 virome sequences
are appearing is very high.  For example, in March 2020, when COVID-19
was declared a pandemic by the world health organization (WHO), there
were a few thousand sequences available.  This grew to tens of
thousands in the late summer, when the Alpha variant emerged in the
UK~\cite{galloway2021emergence}.  By the end of 2020 it was hundreds
of thousands, and in early 2021 it had reached 1 million; today it is
over 2.5 million.  This will likely continue to increase exponentially
(see~\cite{stephens2015genomical}) as many
countries~\cite{cdc_url,bgi_url} ramp up their sequencing
infrastructure for COVID-19 and future pandemics.

\begin{figure}[!ht]
  \centering
  \includegraphics[scale=0.3,page=1]{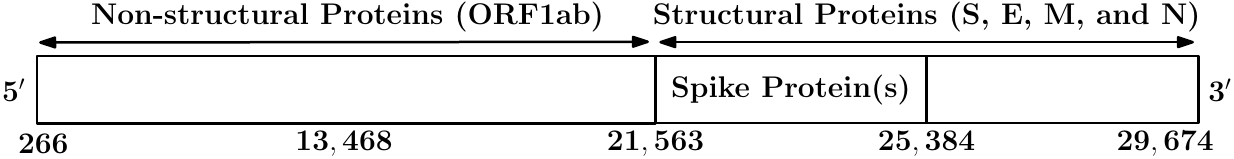}
  \caption{The SARS-CoV-2 genome is composed of ${\approx}30$Kb
    nucleotide base pairs, which codes for several proteins, including
    the spike protein.  The region of the genome which corresponds to
    the spike protein is composed of 3821 ($25,384 - 21,563$)
    nucleotide base pairs, hence $3821$ (+ 1 stop codon *) $/ 3 =
    1274$ amino acids.}
  \label{fig_spike_seq_example}
\end{figure}

The available SARS-CoV-2 virome sequence data, in databases such as
GISAID, has a high \emph{variety}, since it comprises sequences from
all over the world.  Since COVID-19 has spread all over the world for
more than a year now, and viruses continue to mutate over time, there
are quite a number of variants of the SARS-CoV-2 virome, and they
continue to emerge.  Because variants continue to emerge and die off,
some epidemiologists have even proposed a dynamic nomenclature system
similar to that used for the common cold or
flu~\cite{rambaut-2020-nomenclature}.  We use this so-called ``Pango
Lineage'' nomenclature to identify the variants we study here, since
only the very common variants of concern (VoCs) are named.  Examples
of such named VoCs are the Alpha~\cite{galloway2021emergence} (Pango
Lineage B.1.1.7), Gamma~\cite{naveca2021phylogenetic} (P.1) and
Delta~\cite{yadav2021neutralization} (B.1.617.2) variants (see
Table~\ref{tbl_variant_information} for a more complete list).  The
genomic variations (which happen disproportionately in the spike
region, see Figure~\ref{fig_spike_seq_example}) that define these
different variants have been associated with increased
transmissibility~\cite{volz2021assessing}, and immune
evasion~\cite{mccallum2021a}.

Because databases such as GISAID~\cite{gisaid_website_url} collect
sequences from all over the world, they come from heterogeneous
sources of sequencing technologies and centers, leading to multiple
levels of \emph{veracity}.  However, the largest source of different
veracity in the data is the widely varying degree to which different
populations are represented.  For example, the UK sequences about 5\%
of its population of ${\approx}70$ million, the USA sequences about
1\% of its population of ${\approx}300$ million, while India sequences
only a fraction of a percent of its population of ${\approx}1.3$
billion~\cite{FURUSE2021305,maxmen2021why} (see
Figure~\ref{fig_country_wise_distribution}).  Because of this, for
example, even though the Delta variant likely originated in India, the
majority of the available sequences of this variant are from the UK
and the USA, after it arrived in these
countries~\cite{gisaid_website_url}.

Since the genomic sequence of a virus encodes all of its functions
such as virulence and transmissibility, the \emph{value} of such
massive amounts of genomic data is clear.  It is variation in this
genomic sequence itself which defines the different variants of
SARS-CoV-2 such as Alpha, Delta and Gamma.  All of these variants
differ from each other in effect (due to their unique genomic
variations), yet they all descend from the original SARS-CoV-2
sequence~\cite{wu-2020-new}.  It is only through a process of
evolution and transmission to many parts of the globe for over a year,
has it diverged to this extent.  The amount of sequence data available
today puts us in the age of \emph{genomic surveillance}: tracking the
spread of pathogens in terms of genomic
content~\cite{wu2021wastewater,gardy2018surveillance}.

Approaches for rapidly clustering and classifying sequences will be
crucial in these genomic surveillance efforts.  A clustering method,
when applied to the data on a daily basis, for example, would identify
a new and rapidly emerging variant in terms of a cluster which grows
abnormally quickly, allowing scientists to focus on this cluster.
Classification, on the other hand, would allow us to track the spread
of known variants in new municipalities, regions, countries and
continents.  For example, the USA had a wave of the Alpha variant from
the UK in early 2021, and later, a wave of the Delta variant from
India and via other intermediaries, such as the UK (see
Figure~\ref{fig_usa_variant_patterns}).  Such patterns of spread can
reveal information about the underlying transmission networks between
different countries (the UK and USA, or the UK and India), or even
parts of different countries.  This can help overcome some of the
different veracity in the data, such as the widely varying degree to
which different countries are represented in terms of sequencing data,
due to sampling bias.  For example, even though India is very
under-sampled compared to the UK, the wave of the Delta variant in the
UK, along with information about flights from India to the UK near the
beginning of this wave could give us insights on how the Delta variant
originated and spread in India.

The development of clustering and classification approaches needs
several important considerations, however.  For one, the number of
sequences is so huge that any way of extracting useful features
becomes even more critical.  Since the spike protein is the entry
point of the virus to the host cell, it is an important characterizing
feature of a
coronavirus~\cite{li-2016-structure,walls-2020-structure}.  Most of
the variants of SARS-CoV-2 are characterized by mutations which happen
disproportionately in the spike region of the
genome~\cite{galloway2021emergence,naveca2021phylogenetic,yadav2021neutralization}.
Even the mRNA vaccines (\eg, Pfizer and Moderna) for COVID-19 are
designed to target only the SARS-CoV-2 spike
protein~\cite{corbett-2020-mrna} (unlike traditional vaccines which
comprise an entire virome).  Since the spike region is sufficient to
characterize most of the important features of a viral sample, yet is
much smaller in length, as depicted in
Figure~\ref{fig_spike_seq_example}, we focus on an embedding approach
tailored to the spike region of the sequences.

Previously, some efforts have been done to perform classification and
clustering of SARS-CoV-2 spike
sequences~\cite{ali2021k,ali2021effective,kuzmin2020machine}.
However, those methods are not scalable to the amount of data we use
in this study.  Although they were successful in getting high
predictive accuracy, it is not clear if the proposed methods are
robust and will give the same predictive performance on larger
datasets.  In this paper, we propose Spike2Vec, an efficient and
scalable feature vector generation approach for SARS-CoV-2 spike
sequences, to which we can apply different machine learning tasks
downstream, such as classification and clustering.  Our contributions
in this paper are as follows:

\begin{enumerate}
\item We propose an embedding approach, called Spike2Vec that
  outperforms the baseline classification method in terms of
  predictive accuracy.
\item We show that our method is scalable on larger datasets by using
  ${\approx}2.5$ million spike sequences.
\item We prove from the results that the machine learning models used
  in~\cite{ali2021k,ali2021effective,kuzmin2020machine} are not
  scalable on these larger datasets.  This robust checking helps us to
  analyze the machine learning models in detail in terms of their
  appropriateness for SARS-CoV-2 spike sequences.
\item We also show that in terms of clustering, our embedding approach
  is better than the baseline model.
\end{enumerate}

The rest of the paper is organized as follows:
Section~\ref{sec_related_work} contains a discussion on the previous
studies related to our research problem.
Section~\ref{sec_proposed_approach} contains a detailed description of
our Spike2Vec approach.  Section~\ref{sec_exp_setup} contains the
implementation details of the experimental evaluation of Spike2Vec,
along with the dataset statistics and discussion of the baseline
models.  We present and discuss the results of this experimental
evaluation in Section~\ref{sec_results_discussion}.  Finally, we
conclude our paper in Section~\ref{sec_conclusion}.

\section{Literature Review}
\label{sec_related_work}

Because of the rapid spread of COVID-19 since December 2019, volumes
of sequence data are available for this virus.  This new source of
information has attracted researchers from all fields to perform
analysis on this data to better understand the diversity and dynamics
of this virus.  With the number of sequences in the millions, this is
far out of reach for approaches based (at least purely) on
phylogenetic reconstruction, which can handle at most
thousands~\cite{hadfield2018a}, or tens of
thousands~\cite{minh2020iqtree} of sequences.  Researchers have hence
recently turned to clustering and other machine learning (ML)
approaches as an alternative to studying patterns in this data.
Authors in~\cite{kuzmin2020machine} propose a one-hot encoding based
approach to classify different coronavirus hosts using the spike
portion of the virus rather than the entire sequence, obtaining
near-optimal prediction accuracy.  Ali~\ea{} in~\cite{ali2021k}
perform classification of different variants of the human SARS-CoV-2.
Although they were successful in achieving higher accuracy than
in~\cite{kuzmin2020machine}, the kernel method used in their approach,
however, is not scalable to the size of the data we use in this study.
This drawback makes it difficult to use this approach in real-world
scenarios such as the current scenario.

Supervised and unsupervised feature selection methods such as ridge
regression~\cite{ali2021simpler}, lasso regression, and principal
component analysis (PCA)~\cite{ali2019short} are very popular for not
only reducing the runtime, but also for improving the predictive
performance of the underlying machine learning algorithms.  Authors
in~\cite{ali2021effective} perform clustering on SARS-CoV-2 spike
sequences and show that clustering performance could be improved by
simply using lasso and ridge regression.  Although they were also able
to get significant improvement in terms of clustering quality as
compared to the baseline, using feature selection methods like ridge
regression and lasso regression scales very poorly on the larger
datasets, such as the one we use in this study.  Melnyk~\ea,
in~\cite{melnyk2021alpha} perform clustering of the entire SARS-CoV-2
genome (rather than just the spike sequence) using
CliqueSNV~\cite{knyazev2020accurate}, a method originally designed for
identifying haplotypes in an intra-host viral population.  Although
they obtained good overall $F_1$ scores, our (clustering) approach
tends to obtain better overall $F_1$ scores.  It would be interesting
to know whether that is because of our feature vector representation,
or because we leverage more (and more up-to-date) data, or both.

Existing work on fixed length numerical representation of the data
successfully perform different data analytics tasks. It has
applications in different domains such as
graphs~\cite{hassan2020estimating,Hassan2021Computing}, nodes in
graphs~\cite{ali2021predicting,grover2016node2vec}, and electricity
consumption~\cite{ali2019short,Ali2020ShortTerm}.  This vector-based
representation also achieves significant success in sequence analysis,
such as
texts~\cite{shakeel2020multiDataAugmentation,Shakeel2020LanguageIndependent,shakeel2019multiBilingualSmsClassification},
electroencephalography and electromyography
sequences~\cite{atzori2014electromyography,ullah2020effect},
networks~\cite{Ali2019Detecting}, and biological
sequences~\cite{ali2021simpler,Kuksa_SequenceKernel}.  However, most
of the existing sequence classification methods require the input
sequences to be aligned. Although sequence alignment help to analyze
the data better, it is a very costly process.  Several efforts have
been made previously to identify the transmission patterns of
different variants that can help the appropriate authorities to devise
appropriate public health interventions so that the rapid spread of
viruses can be avoided
\cite{Ahmad2016AusDM,ahmad2017spectral,Tariq2017Scalable,AHMAD2020Combinatorial}.

Farhan~\ea, in~\cite{farhan2017efficient} propose an efficient
approach to compute a similarity matrix (kernel matrix).  The computed
kernel matrix is proven to be efficient for sequence classification.
However, since their approach requires to save an entire $n \times n$
dimensional kernel matrix (where $n$ is the total number of
sequences), this makes their method expensive in terms of space.
Authors in~\cite{rahimi2007random} use the random feature method to
map the original input into a low dimensional feature space so that
the inner product of the low dimensional data is approximately equal
to the inner product of the original data points.  Peng~\ea,
in~\cite{peng2021random} use the random feature attention model for
text classification.  Their approach is linear in terms of runtime and
space, and uses random feature methods to approximate the softmax
function.

While dealing with \emph{Big Data}, it is important to analyze the
trade-off between the prediction accuracy and the
runtime~\cite{ali2021cache}.  Although Ali~\ea, in~\cite{ali2021k} use
the kernel method for spike sequence classification, since the kernel
computation is, however, expensive in terms of time and space, their
approach is only a proof of concept, and not feasible in a real-world
scenario.

\section{Proposed Approach}
\label{sec_proposed_approach}

In this section, we give a step by step description of the Spike2Vec
approach.  From the SARS-CoV-2 spike sequences, we first generate
$k$-mers so that we can preserve some ordering information of the
sequences.  It is interesting to note that \emph{de novo} genome
assembly (when no reference is present) involves inferring sequence
order by assembling $k$-mers obtained from short
reads~\cite{rizzi-2019-overlap}.  After the $k$-mers are generated, to
convert the alphabetical information of $k$-mers into a fixed length
numerical representation (so that ML algorithms can be applied), we
generate frequency vectors, which count the number of occurrences of
each $k$-mer in the spike sequence.  We then map the high dimensional
frequency vectors to a low dimensional embedding using an approximate
kernel approach.  Each step of Spike2Vec is explained in detail below.

\subsection{$k$-mers Generation}

\begin{figure}[!ht]
  \centering
  \includegraphics[scale=0.25]{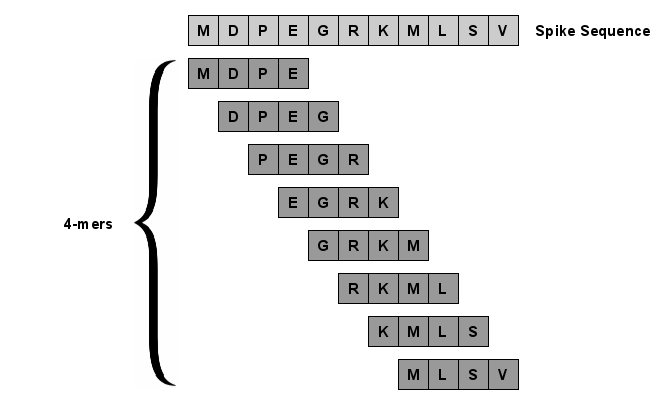}
  \caption{Example of 4-mers of the amino acid sequence
    ``MDPEGRKMLSV''.}
  \label{fig_k_mer_demo}
\end{figure}

The first step of Spike2Vec is to compute all $k$-mers for the spike
sequences. The main idea behind using $k$-mers is to allow some
ordering information of the sequence to be preserved.  The total
number of $k$-mers, which we can be generated from a spike sequence of
length $N$ is $N - k + 1$, where $N = 1274$ and $k$ is a user-defined
parameter for the size of each mer.  See Figure~\ref{fig_k_mer_demo}
for an example.  In our experiments we use $k = 3$; this was decided
using a standard validation set approach~\cite{validationSetApproach}.

\subsection{Frequency Vectors Generation}

Since $k$-mers is an alphabets-based representation of a spike
sequence, we need to convert the $k$-mers into a numerical
representation.  Therefore, we design a feature vector that contains
the count of each $k$-mer in its respective spike sequence. Each
sequence $A$ is over an alphabet $\Sigma$.  Note that alphabets in our
dataset represent amino acids of the spike sequence.  These fixed
length frequency vectors have length $|\Sigma|^k$ --- the number of
possible $k$-mers of a spike sequence.  Since the total number of
alphabets in our data are $21$ (the number of amino acids), the length
of each frequency vector becomes $21^{3} = 9261$.

\subsection{Low Dimensional Representation}

In large scale machine learning (ML) tasks such as classification and
clustering, typical supervised and unsupervised dimensionality
reduction methods such as principal component analysis, ridge
regression, and lasso regression, etc., are not suitable because they
take a lot more time to execute. Therefore, in a real world scenario
where we can have a huge amount of data, the scalability of any
underlying algorithm could be one of the major issues. One option is
to use kernel based algorithms that compute a similarity matrix which
can later be used for the underlying ML tasks. To compute the kernel
matrix (gram matrix), the kernel trick is used.
\begin{definition}[Kernel Trick]
  The Kernel Trick is used to generate features for an algorithm which
  depends on the inner product between only the pairs of input data
  points. The main idea is to avoid the need to map the input data
  (explicitly) to a high-dimensional feature space.
\end{definition}
The Kernel Trick relies on the following observation: \textit{Any
  positive definite function f(a,b), where $a,b \in \mathcal{R}^d$,
  defines an inner product and a lifting $\phi$ so that we can quickly
  compute the inner product between the lifted data
  points}~\cite{rahimi2007random}. More formally:
\begin{equation}
  \langle \phi (a), \phi (b) \rangle = f(a,b)
\end{equation}
The major drawback of kernel methods is that in case of large training
data, the kernel method suffers from large initial computational and
storage costs.

To overcome these computational and storage problems, we use an
approximate kernel method called random Fourier features
(RFF)~\cite{rahimi2007random}, which maps the input data to a
randomized low dimensional feature space (euclidean inner product
space). More formally:
\begin{equation}
  z: \mathcal{R}^d \rightarrow \mathcal{R}^D
\end{equation}
In this way, we approximate the inner product between a pair of
transformed points. More formally:
\begin{equation}\label{eq_z_value}
  f(a,b) = \langle \phi (a), \phi (b) \rangle \approx z(a)' z(b)
\end{equation}
In Equation~\eqref{eq_z_value}, $z$ is low dimensional (unlike the
lifting $\phi$). In this way, we can transform the original input data
with $z$, which acts as the approximate low dimensional embedding for
the original data. This low dimensional representation is then used as
an input for different ML tasks like classification and regression.

\section{Experimental Evaluation}
\label{sec_exp_setup}

We now detail the experiments we performed to evaluate Spike2Vec in
terms of both the downstream classification and clustering results
obtained.

\subsection{Experimental Setup}

All experiments are conducted using an Intel(R) Xeon(R) CPU E7-4850 v4
@ $2.10$GHz having Ubuntu $64$ bit OS ($16.04.7$ LTS Xenial Xerus)
with 3023 GB memory.  Implementation of Spike2Vec is done in Python
and the code is available online for
reproducibility\footnote{\url{https://github.com/sarwanpasha/Spike2Vec}}. Our
pre-processed data is also available
online\footnote{\url{https://drive.google.com/drive/folders/1-YmIM8ipFpj-glr9hSF3t6VuofrpgWUa?usp=sharing}},
which can be used after agreeing to terms and conditions of
GISAID\cite{gisaid_website_url}.  For the classification algorithms,
we use $10\%$ of the data for training and $90\%$ for testing. The
purpose of using smaller training dataset is to show how much
performance gain we can achieve while using minimal training
data. Note that our data split and pre-processing follow those
of~\cite{ali2021k}.

\subsection{Dataset Statistics}

We used the (aligned) amino acid sequences corresponding to the spike
protein from the largest known database of SARS-CoV-2 sequences,
GISAID~\cite{gisaid_website_url}.  In our dataset, we have 2,519,386
spike sequences along with the COVID-19 variant information (in our
data, we have $1327$ variants in total) for each spike sequence. The
information about some of the more well-represented variants is given
in Table~\ref{tbl_variant_information}. Since most of the variants are
new, we do not have all the information available for all
them. Therefore, we put ``-" in any field of
Table~\ref{tbl_variant_information} for which we do not have any
information available.

\begin{table}[ht!]
  \centering
  \begin{tabular}{p{0.9cm}lp{0.7cm}p{2cm} | p{1.2cm}}
    \hline
    Pango Lin. & \multirow{2}{*}{Region} & \multirow{2}{*}{Labels} &
    \multirow{2}{*}{No. Mut. S/Gen.} &  No. sequences\\
    \hline	\hline	
    B.1.1.7 & UK~\cite{galloway2021emergence} &  Alpha & 8/17 & \hskip.1in 976077\\
    B.1.351  & South Africa~\cite{galloway2021emergence}  &  Beta & 9/21& \hskip.1in 20829\\
    B.1.617.2  & India~\cite{yadav2021neutralization}  &  Delta &  8/17  & \hskip.1in 242820\\
    P.1  &  Brazil~\cite{naveca2021phylogenetic} &  Gamma &  10/21 & \hskip.1in 56948\\
    B.1.427   & California~\cite{zhang2021emergence}  & Epsilon  &  3/5 & \hskip.1in 17799\\
    AY.4   & India~\cite{CDS_variantDef}  & Delta  &  - & \hskip.1in 156038\\
    B.1.2   & -  & -  & -  & \hskip.1in 96253\\
    B.1   &  & & & \hskip.1in 78741 \\
    B.1.177   &  - &  - & -  & \hskip.1in 72298\\
    B.1.1   & -  & & -  & \hskip.1in 44851 \\
    B.1.429   &  - &  - & -  & \hskip.1in 38117\\
    AY.12   & India~\cite{CDS_variantDef}  & Delta  & -  & \hskip.1in 28845\\
    B.1.160   & -  & -  &  - & \hskip.1in 25579\\
    B.1.526   & New York \cite{west2021detection}  &  Iota & 6/16 & \hskip.1in 25142\\
    B.1.1.519   & -  & -  & -  & \hskip.1in 22509\\
    B.1.1.214   & -  & -  &  - & \hskip.1in 17880\\
    B.1.221   & -  & -  & -  & \hskip.1in 13121\\
    B.1.258   & -  & -  & -  & \hskip.1in 13027\\
    B.1.177.21   & -  &  - & -  & \hskip.1in 13019\\
    D.2   &  - &  - &  - & \hskip.1in 12758\\
    B.1.243   & -  & -  & -  & \hskip.1in 12510\\
    R.1   & -  & -  & -  & \hskip.1in 10034\\
    \hline
  \end{tabular}
  \caption{The SARS-CoV-2 variants which were represented in more than
    10,000 sequences (of the ${\approx}2.5$ million sequences). The
    S/Gen. column represents the number of mutations in the Spike (S)
    region / entire genome.  The total number of amino acid sequences
    in our dataset is 2,519,386. The variants listed in this table
    comprise 1,995,195 sequences.}
  \label{tbl_variant_information}
\end{table}

\begin{figure}[ht!]
  \centering
  \includegraphics[scale=0.3]{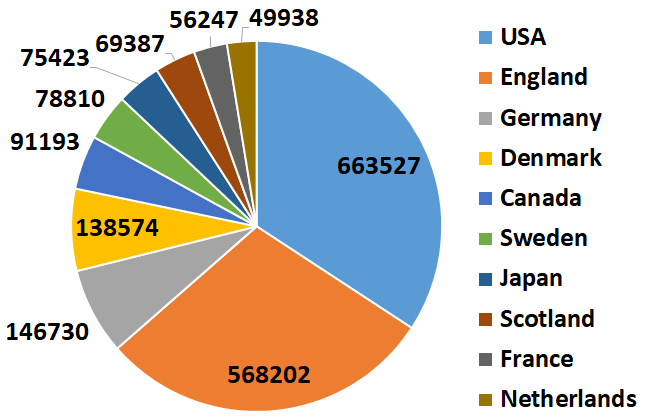}
  \caption{Country-wise distribution (for the top 10 countries) of
    spike sequences.}
  \label{fig_country_wise_distribution}
\end{figure}

Figure~\ref{fig_country_wise_distribution} shows the total number of
spike sequences for the top $10$ countries worldwide. In our GISAID
dataset, a total of $219$ countries are represented. Since the USA has
the highest number of spike sequences, we use it in a case study to
analyze the spread patterns of different variants in
Section~\ref{sec_usa_case_study}.

\subsection{Data Visualization}

\begin{figure}[!ht]
  \centering
  \includegraphics[scale=0.3]{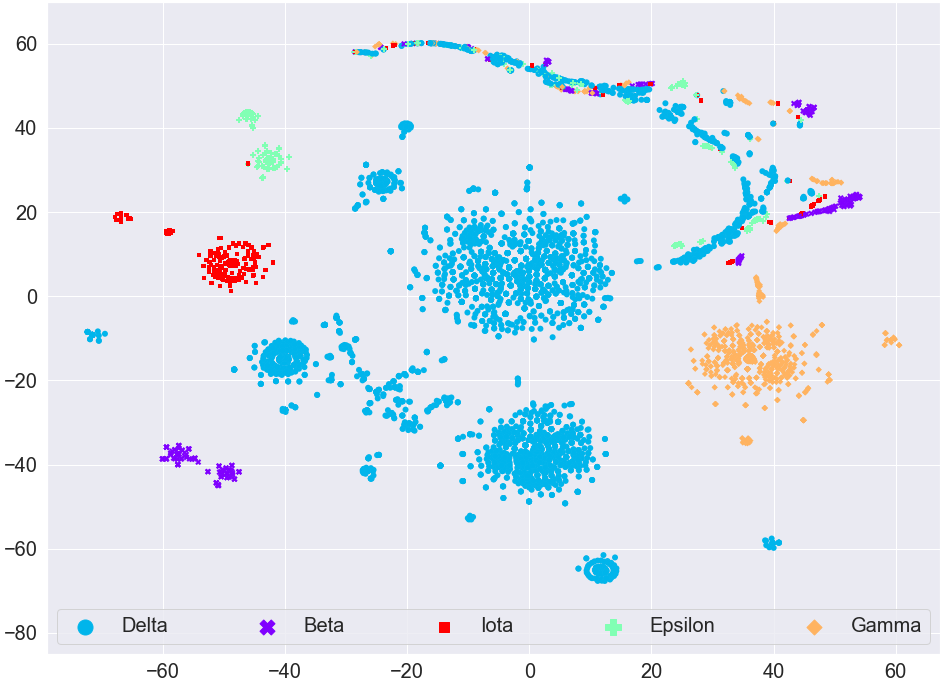}
  \caption{t-SNE embeddings of spike sequences}
  \label{fig_tsne_plot}
\end{figure}

To see if there is any (hidden) clustering in the data, we mapped the
data to 2D real vectors using the t-distributed stochastic neighbor
embedding (t-SNE) approach~\cite{van2008visualizing}.  Since it was
not possible to run the t-SNE algorithm on all ${\approx}2.5$ million
spike sequences, we obtained a representative subset of sequences
containing 7000 randomly selected sequences such that the proportion
of each variant in this subset is equal to its proportion in the
original data.  The t-SNE plot for Delta, Beta, Iota, Epsilon, and
Gamma variants is shown in Figure~\ref{fig_tsne_plot}.

\subsubsection{USA Case Study}
\label{sec_usa_case_study}

\begin{figure}[h!]
  \centering
  \includegraphics[scale=0.25]{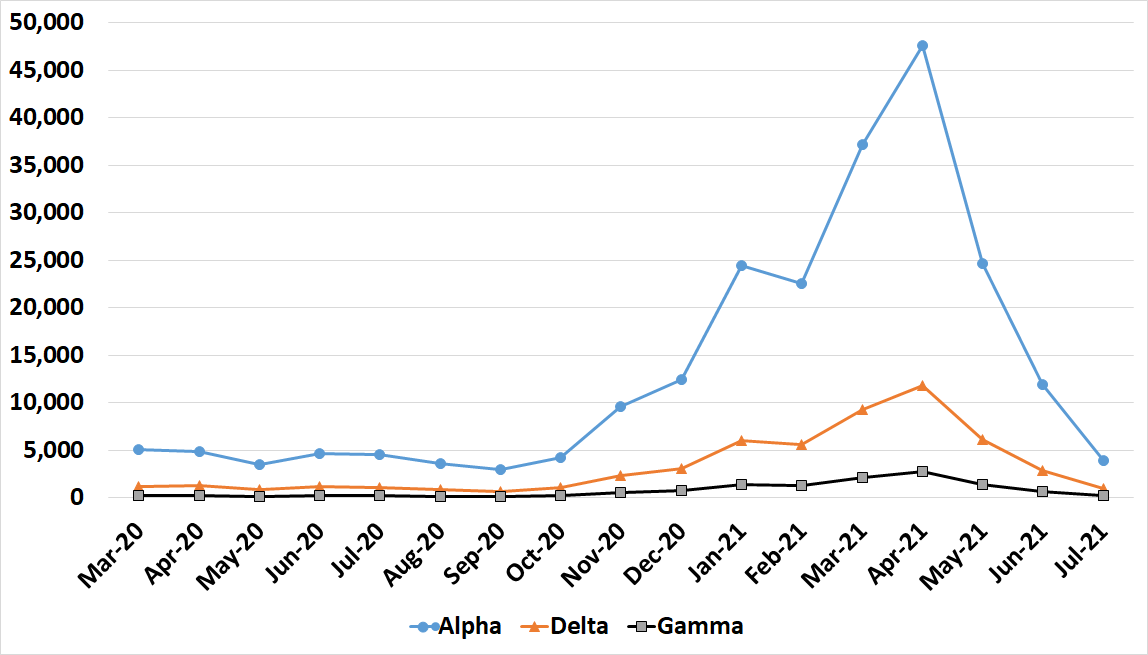}
  \caption{Spread pattern of Alpha (blue line), Delta (orange line),
    and Gamma (black line) variants in the USA from March 2020 to July
    2021. The y-axis shows the total number of COVID-19 infected
    patients.}
  \label{fig_usa_variant_patterns}
\end{figure}

Figure~\ref{fig_usa_variant_patterns} shows the COVID-19 spread
pattern for three variants in the USA from March 2020 to July 2021.
We can see in Figure~\ref{fig_usa_variant_patterns} that after the
coronavirus spread hit its peak in April $2021$, the number of cases
of the coronavirus started decreasing. That was the point where a
significant proportion of the population of the USA was vaccinated
(hence peak spread reduced).

To evaluate Spike2Vec, we perform classification and clustering on the
low dimensional feature vectors that it produces. For the
classification task, we use naive Bayes (NB), logistic regression
(LR), and ridge classifier (RC)~\cite{singh2016comparison}. For the
clustering analysis, we use the $k$-means algorithm. We report
accuracy, precision, recall, weighted $F_1$, macro $F_1$, and ROC-AUC
for classification and weighted $F_1$ score (with the major variant in
a cluster as the label) for clustering evaluation. For $k$-means, we
use $22$ as the number of clusters. We selected this number of
clusters using the elbow method~\cite{ali2021effective}. For this, we
perform clustering with different numbers of clusters from $2$ to
$100$ and then see the trade-off between the \emph{sum of squared
  error} (distortion score) and the runtime. After analyzing this
trade-off, we use the knee point detection algorithm
(KPDA)~\cite{satopaa2011finding} to find the optimal value of $k$ for
the $k$-means.

\begin{table*}[ht!]
  \centering
  \begin{tabular}{p{0.7cm}p{0.3cm}p{0.09cm}p{0.55cm}p{0.4cm}p{0.4cm}p{0.5cm}p{0.3cm}p{0.09cm}p{0.3cm}p{0.3cm}p{0.3cm}p{0.3cm}p{0.3cm}p{0.3cm}p{0.3cm}p{0.4cm}p{0.3cm}p{0.6cm}p{0.5cm}p{0.3cm}p{0.3cm}p{0.3cm}}
    \hline
    & \multicolumn{22}{c}{$k$-means (Cluster IDs)} \\
    \cline{2-23}
    Variant & 0 & 1 & 2 & 3 & 4 & 5 & 6 & 7 & 8 & 9 & 10 & 11 & 12 & 13 & 14 & 15 & 16 & 17 & 18 & 19 & 20 & 21 \\
    \hline \hline
    Epsilon & 109 & 0 & 3186 & 432 & 113 & 219 & 67 & 0 & 160 & 9 & 134 & 0 & 18 & 14 & 0 & 78 & 9 & 3792 & 113 & 0 & 48 & 41 \\
    Alpha & 6061 & 1 & 175923 & 23353 & 5846 & 11754 & 3376 & 0 & 9466 & 1041 & 6889 & 0 & 734 & 1281 & 0 & 4160 & 329 & 205848 & 5730 & 0 & 3193 & 2136 \\
    Gamma & 344 & 0 & 10403 & 1312 & 327 & 686 & 205 & 0 & 534 & 63 & 390 & 0 & 52 & 77 & 0 & 254 & 14 & 11977 & 324 & 0 & 182 & 137 \\
    Beta & 144 & 0 & 3853 & 436 & 115 & 237 & 64 & 0 & 191 & 8 & 148 & 0 & 19 & 25 & 0 & 81 & 7 & 4435 & 119 & 0 & 71 & 44 \\
    Delta & 1432 & 1 & 43691 & 5732 & 1391 & 2832 & 831 & 0 & 2342 & 241 & 1777 & 0 & 172 & 315 & 0 & 1016 & 77 & 51596 & 1400 & 0 & 836 & 541 \\    
    \hline
  \end{tabular}
  \caption{Contingency tables of variants vs. clusters.}
  \label{tbl_contingency}
\end{table*}

\subsection{Baseline Algorithm}

As a baseline, we use the one-hot encoding (OHE)
method~\cite{kuzmin2020machine}.  In spike sequences, we have $21$
unique amino acids (unique alphabets forming $\Sigma$) namely
``ACDEFGHIKLMNPQRSTVWXY''. Also, the length of each spike sequence is
$1273$ plus a terminating character $*$ at the $1274^{th}$
location. After getting the OHE for each spike sequence, we get a
feature vector of length $26,733$ corresponding to each spike sequence
($21 \times 1273 = 26,733$).  We then use RFF on the OHEs to get the
low dimensional feature vector representation (to avoid the curse of
dimensionality).

\section{Results and Discussion}
\label{sec_results_discussion}

In this section, we present the classification and clustering results
for Spike2Vec and its comparison with OHE and report the importance of
each amino acid using information gain.

\subsection{Classification Results}

Results for different classification algorithms are shown in
Table~\ref{tble_classification_results}. Note that overall logistic
regression is a clear winner in case of Spike2Vec. All of the
classifiers in case of Spike2Vec clearly outperform the corresponding
classifiers with OHE. This performance for different evaluation
metrics shows the effectiveness of using $k$-mers instead of OHE for
representing the spike sequences. Also, we can observe that although
the performance of RC in the case of Spike2Vec is not better than LR,
it is significantly better than LR and NB, however, in terms of
training runtime. Therefore, we can conclude that overall LR is better
in terms of prediction performance with Spike2Vec while RC is better
in terms of runtime along with comparable performance to LR.
\begin{table}[!ht]
  \centering
  \begin{tabular}{p{0.9cm}p{0.3cm}p{0.4cm}p{0.4cm}p{0.4cm}p{0.7cm}p{0.65cm}p{0.5cm} | p{0.8cm}}
    \hline
    \multirow{3}{*}{Approach} & \multirow{3}{0.3cm}{ML Algo.} & \multirow{3}{*}{Acc.} & \multirow{3}{*}{Prec.} & \multirow{3}{0.4cm}{Recall} & \multirow{3}{0.7cm}{$F_1$ (Weig.)} & \multirow{3}{0.65cm}{$F_1$ (Macro)} & \multirow{3}{0.5cm}{ROC-AUC} & Training time (sec.) \\	
    \hline	\hline	
    \multirow{3}{*}{OHE}  
    & NB & 0.30 & 0.58 & 0.30 & 0.38 & 0.17 & 0.59  &  566.09\\
    & LR & 0.56 & 0.49 & 0.56 & 0.49 & 0.19 & 0.57 & 1309.06\\
    & RC & 0.56 & 0.47 & 0.56 & 0.48 & 0.17 & 0.56  & 110.76\\
    \hline
    \multirow{3}{*}{Spike2Vec}  
    & NB & 0.42 & \textbf{0.79} & 0.42 & 0.52 & 0.39 & 0.68  &  457.54\\
    & LR & \textbf{0.68} & 0.68 & \textbf{0.68} & \textbf{0.64} & \textbf{0.49} &  \textbf{0.69}  & 830.63\\
    & RC & 0.67 & 0.68 & 0.67 &  0.62 & 0.44 & 0.67  & \textbf{95.73}\\
    \hline
  \end{tabular}
  \caption{Variants Classification Results (10\% training set and 90\%
    testing set) for the top $22$ variants ($1995195$ spike sequences)
    listed in Table~\ref{tbl_variant_information}. Best values are
    shown in bold.}
  \label{tble_classification_results}
\end{table}

\subsection{Clustering Results}

We also test the performance of Spike2Vec using the $k$-means
clustering method.  The contingency table for $k$-means (for some of
the more well represented variants) computed using Spike2Vec is given
in Table~\ref{tbl_contingency}. Since we have a total of $1327$
variants, it is not possible to show results for all
variants. Therefore, we only present results for some of the common
variants.  The weighted $F_1$ score for OHE and Spike2Vec are shown in
Table~\ref{tbl_f1_weighted_clustering}. We can observe that Spike2Vec
clearly outperforms OHE in case of all but one variant. The reason for
the poor performance in the case of Beta and Epsilon variants is due
to the fact that they are in comparatively smaller proportion in the
dataset (see Table~\ref{tbl_variant_information}). Because of this,
Spike2Vec is not able to design a rich feature vector representation
of these variants.

\begin{table}[ht!]
  \centering
  \begin{tabular}{lccccc}
    \hline
    & \multicolumn{5}{c}{$F_1$ Score (Weighted) for Different Variants} \\
    \cline{2-6}
    Methods & Alpha & Beta & Delta &  Gamma & Epsilon \\
    \hline	\hline	
    OHE & 0.0410 & \textbf{0.0479} & 0.5942 & 0.6432 & 0.0571 \\
    Spike2Vec & \textbf{0.9997} & 0.0300 & \textbf{0.8531} & \textbf{0.9680} & \textbf{0.2246} \\
    \hline
  \end{tabular}
  \caption{$F_1$ scores for five variants from the $k$-means
    clustering algorithm on all $1327$ variants (2519386 spike
    sequences) in the GISAID dataset. Best values are in bold.}
  \label{tbl_f1_weighted_clustering}
\end{table}

\subsection{Importance of each Amino Acid}

\begin{figure}[ht!]
  \centering
  \includegraphics{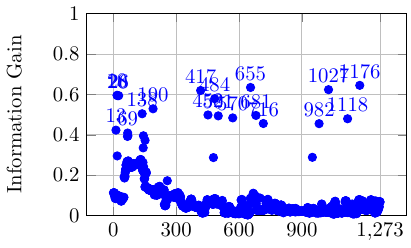}
  \caption{Information gain for each amino acid position (x-axis) with
    respect to the variants.}
  \label{fig_data_correlation}
\end{figure}

To evaluate the importance of amino acid $a$ in a set $V$ of variants,
we compute the information gain $\mbox{IG}(V,a) = H(V) - H(V | a)$,
where $H = \sum_{v \in V} -p_v \log p_v$. Note that $H$ is the
entropy, and $p_v$ is the probability of the class $v \in V$.  We
extracted a sample dataset of $20000$ spike sequences and computed IG
(see Figure~\ref{fig_data_correlation}).  The USA based Centers for
Disease Control and Prevention (CDC) catalogs mutations that take
place in different variants~\cite{CDS_variantDef}. We compare the
mutation information from the CDC with our (high) IG values. According
to the CDC, R452L is present in Epsilon and Delta lineages and
sub-lineages while K417N, E484K, and N501Y substitutions are present
in the Beta variant.  Note that R452L means that amino acid at
position 452 mutated from `R' to `L'.  Similarly, K417T, E484K, and
N501Y substitutions are present in the Gamma
variant~\cite{CDS_variantDef}.  We can see in
Figure~\ref{fig_data_correlation} that we obtained the maximum IG
values for many of the same amino acids positions mentioned by CDC.

\section{Conclusion}
\label{sec_conclusion}

We propose an embedding approach that can be used to perform different
machine learning tasks on the SARS-CoV-2 spike sequences. We show that
our model can scale to several million sequences, and it also
outperforms the baseline models significantly. Since the COVID-19
disease is relatively new, we do not have enough information available
for different coronavirus variants so far. We will explore the new
(and existing) variants in more detail in the future. We will also use
deep learning models to enhance the prediction performance of
Spike2Vec.

\subsection{Acknowledgments}

This research was supported by a Georgia State University startup
grant.

\bibliographystyle{IEEEtran.bst}
\bibliography{spike_to_vec}

\end{document}